\documentclass[fleqn,10pt]{wlscirep}

\usepackage{soul}
\usepackage{longtable}
\usepackage{bm}
\usepackage{dcolumn}
\usepackage{epsfig}
\usepackage{graphicx}
\usepackage{subfigure}
\usepackage{comment}
\usepackage{amsfonts}
\usepackage{amsmath}
\usepackage{amssymb}
\usepackage{subfigure}

\usepackage{color}

\newcommand{\be}{\begin{equation}}
\newcommand{\ee}{\end{equation}}
\newcommand{\bea}{\begin{eqnarray}}
\newcommand{\eea}{\end{eqnarray}}

\def\vec{\mathbf}
\def\mc{\mathcal}

\usepackage{times}
\usepackage{appendix}

\begin{document}

\title{The vicinity of hyper-honeycomb $\beta$-Li$_2$IrO$_3$ to a three-dimensional Kitaev spin liquid state}

\author[1,*]{Vamshi M.~Katukuri}
\author[1]{Ravi Yadav}
\author[1]{Liviu Hozoi}
\author[1,2,+]{Satoshi Nishimoto}
\author[1,3]{Jeroen van den Brink}
\affil[1]{Institute for Theoretical Solid State Physics, IFW Dresden, Helmholtzstr.~20, 01069 Dresden, Germany}
\affil[2]{Institute for Theoretical Physics, Technische Universit\"{a}t Dresden, Helmholtzstr.~10, 01069 Dresden, Germany}
\affil[3]{Department of Physics, Harvard University, Cambridge, Massachusetts 02138, USA}
\affil[*]{v.m.katukuri@ifw-dresden.de}
\affil[+]{s.nishimoto@ifw-dresden.de}

\begin{abstract}
Due to the combination of a substantial spin-orbit coupling and correlation effects, iridium oxides hold a prominent place in the search for novel 
quantum states of matter, including, e.g., Kitaev spin liquids and topological Weyl states.
We establish the promise of the very recently synthesized hyper-honeycomb iridate $\beta$-Li$_2$IrO$_3$ in this regard.
A detailed theoretical analysis   
reveals the presence of large ferromagnetic first-neighbor Kitaev interactions, while a second-neighbor antiferromagnetic Heisenberg exchange drives 
the ground state from ferro to zigzag order via a three-dimensional Kitaev spin liquid and an incommensurate phase.
Experiment puts the system in the latter regime but the Kitaev spin liquid is very close and reachable by a slight modification
of the ratio between the second- and first-neighbor couplings, for instance via strain.
\end{abstract}

\date\today
\maketitle

\section*{Introduction} 
\begin{figure}[!ht]
\includegraphics[width=1.0\linewidth]{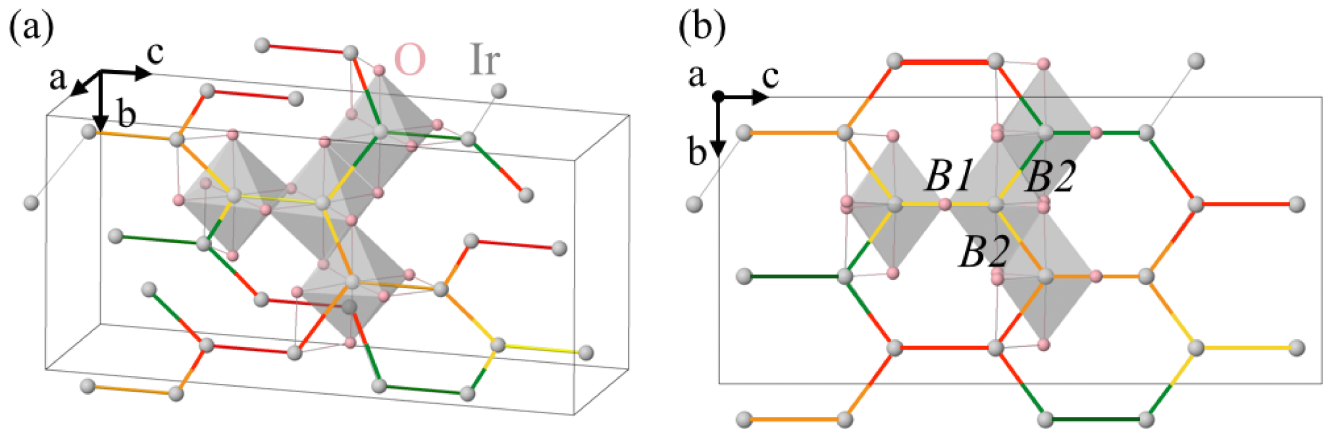}
\caption{
(a) Ir hyper-honeycomb lattice of $\beta$-Li$_2$IrO$_3$. 
The Ir-Ir links along the $c$ axis, associated with equilateral Ir$_2$O$_2$ plaquettes \cite{Takayama14}
and labeled {\it B1}, are shown in four different colors.
{\it B1} links located at 0.125, 0.375, 0.625
 and 0.875 on the $a$ axis are shown in green, yellow,
orange
 and red color, respectively.
{\it B2} bonds connecting the {\it B1} links are shown in dual colors.
O ions around four of the Ir sites are also shown.
(b) Projection of the unit cell on the $bc$ plane. 
}
\label{hyperhoneycomb_lattice}
\end{figure}

In magnetism, frustration refers to the existence of competing exchange interactions that cannot be
simultaneously satisfied.
Such effects can spawn new states of matter with quite exotic physical properties.
Most famous in this regard are the different kinds of quantum spin liquids (QSL's) that emerge from
frustrated spin couplings~\cite{Balents10}.
In these collective states of matter quantum fluctuations are so strong that they disorder the spins
even at the lowest temperatures.
The types of QSL states that then emerge range from chiral ones~\cite{Messio12,Capponi13} to $Z_2$
topological spin liquids~\cite{Misguich02,Yan11,Depenbrock12} carrying fractionalized excitations.
Both experimentally and theoretically such QSL's have been observed and intensely studied in two-dimensional
(2D) systems~\cite{Balents10,Messio12,Capponi13,Misguich02,Yan11,Depenbrock12,Mendels07,Han12,Clark13}.
How this situation carries over to three spatial dimensions (3D), in which tendencies towards formation of
long-range ordered magnetic states are in principle stronger and the disordering effect of quantum
fluctuations therefore less potent, is largely unexplored.
This is not only due to the limitations of theoretical and numerical approaches in 3D but also to the
the sparsity of relevant candidate materials~\cite{Okamoto07}.
Very recently the latter however fundamentally changed through the synthesis of insulating Li$_2$IrO$_3$
polymorphs~\cite{Takayama14,Modic14,Kimchi14} in which the magnetic moments of Ir$^{4+}$ ions form 3D
honeycomb structures with threefold coordination.
Here we concentrate on the $\beta$-Li$_2$IrO$_3$ polymorph, which forms a so-called hyper-honeycomb
lattice, see Fig.~\ref{hyperhoneycomb_lattice}.
Such a lattice might in principle support a 3D Kitaev spin liquid~\cite{Mandal09,ELee14,SBLee14,Nasu14},
a direct counterpart of its lower-dimensional, 2D equivalent~\cite{Kit_kitaev_06,Ir213_KH_jackeli_09,Ir213_KH_chaloupka_10}.

The 2D Kitaev-Heisenberg model on the honeycomb lattice is characterised by the presence of large uniaxial
symmetric magnetic couplings that cyclically permute on the bonds of a given hexagonal ring \cite{Kit_kitaev_06,Ir213_KH_jackeli_09,Ir213_KH_chaloupka_10}.
A QSL phase is present in this model if the ratio between the Kitaev interaction $K$ and Heisenberg
coupling $J$ is larger than 8~\cite{Ir213_KH_chaloupka_10}.
Quasi-2D honeycomb compounds initially put forward for the experimental realization of the Kitaev-Heisenberg
Hamiltonian are $5d^5$ and $4d^5$ $j\!\approx\!1/2$ systems \cite{Ir213_KH_jackeli_09,book_abragam_bleaney}
such as Na$_2$IrO$_3$, $\alpha$-Li$_2$IrO$_3$ and Li$_2$RhO$_3$.
Subsequent measurements evidenced, however, either antiferromagnetically ordered 
\cite{Ir213_jkj2j3_singh_2012,Ir213_choi_2012,Ir213_ye_2012,Ir213_zigzag_liu_2011} or spin-glass
\cite{Li2RhO3_HK_luo_13} ground states in these materials.

The three factors that complicate a straightforward materialisation of the Kitaev QSL ground state in
the quasi-2D honeycomb compounds are the presence of
(i) appreciable additional exchange anisotropies \cite{Na2IrO3_vmk_14,Kee_PRL_2014,Ir213_yamaji_2014},
(ii) two crystallographically inequivalent Ir-Ir bonds and
(iii) longer-range magnetic interactions between second- and third-neighbor iridium moments
\cite{Ir213_jkj2j3_singh_2012,Ir213_jkj2j3_kimchi_2011,Ir213_choi_2012,Ir213_KH_mazin_2013,Na2IrO3_vmk_14}.
These additional interactions push quasi-2D Na$_2$IrO$_3$ and $\alpha$-Li$_2$IrO$_3$ towards the formation
of long-range antiferromagnetic (AF) order at temperatures below 15 K.
Also the 3D honeycomb system $\beta$-Li$_2$IrO$_3$ orders magnetically: at 38 K the spins form an
incommensurate (IC) ordering pattern \cite{biffin_xray} with strong ferromagnetic (FM) correlations
\cite{Takayama14}.
Apparently additional interactions beyond only the nearest-neighbor (NN) Kitaev and Heisenberg ones
are relevant also in the 3D system.
This leaves two main challenges: first, one would like to precisely quantify the different magnetic
exchange interactions between the Ir moments and second, one should like to determine how far away the
magnetic ground state is from a Kitaev-type 3D QSL.
Here we meet these challenges through a combination of {\it ab initio} quantum chemistry calculations
by which we determine the NN magnetic couplings in $\beta$-Li$_2$IrO$_3$ and exact diagonalization (ED)
of the resulting effective spin Hamiltonian, on large clusters, to determine how far
$\beta$-Li$_2$IrO$_3$ is situated from the QSL ground state in the magnetic phase diagram.

The {\it ab initio} results show that the NN exchange in $\beta$-Li$_2$IrO$_3$ is mostly FM, with 
relatively weak FM Heisenberg couplings of a few meV, large FM Kitaev interactions in the range of 
10--15 meV, and additional anisotropies not included in the plain Kitaev-Heisenberg model.
The sign and magnitude of second-neighbor Heisenberg couplings we determine from fits of the ED
calculations to the experimental magnetization data.
This second-neighbor effective coupling comes out as $J_2\!\approx\!0.2\!-\!0.3$ meV and is thus small
and AF.
Remarkably, this AF $J_2$ stabilizes an IC magnetic structure that puts the system to be only a jot 
apart from the transition to a QSL ground state.
Our findings provide strong theoretical motivation for further investigations on the material
preparation side.
The Kitaev QSL phase might be achieved by for instance epitaxial strain and relaxation in
$\beta$-Li$_2$IrO$_3$ thin films, slightly modifying the $J_2/K$ ratio.


\section{Results}
\subsection{Quantum Chemistry Calculations}
Quantum chemistry 
calculations were first performed for the on-site $d$-$d$ excitations, on embedded clusters
consisting of one central octahedron and the three adjacent octahedra (for technical details, see
Supplementary Information (SI) and Ref.\,\citeonline{Ir213_rixs_gretarsson_2013}).
Reference complete-active-space (CAS) multiconfigurational wave functions \cite{book_QC_00}
were in this case generated with an active orbital space defined by the five $5d$ functions
at the central Ir site.
While all possible occupations are allowed within the set of Ir $5d$ orbitals, double occupancy
is imposed in the CAS calculations on the O $2p$ levels and other lower-energy orbitals.
The self-consistent optimization was here carried out for an average of four states, i.e., $^2T_{2g}$
($t_{2g}^5$) and the states of maximum spin multiplicity associated with each of the $t_{2g}^4e_g^1$
and $t_{2g}^3e_g^2$ configurations.
We then subsequently performed multireference configuration-interaction (MRCI) calculations
\cite{book_QC_00} with single and double excitations out of the Ir $5d$ and O $2p$ shells at the
central octahedron.
MRCI relative energies, without and with spin-orbit coupling (SOC), are listed in Table\,I. 

Due to slight distortion of the O cage \cite{Takayama14} and possibly anisotropic
fields associated with the extended surroundings, the degeneracy of the Ir $t_{2g}$ levels is
lifted.
Without SOC, the Ir $t_{2g}^5$ states are spread over an energy window of $\approx$0.1 eV
(see Table\,\ref{dd_exc}).
Similar results were earlier reported for the quasi-2D honeycomb iridates \cite{Ir213_rixs_gretarsson_2013}.
The low-symmetry fields additionaly remove the degeneracy of the $j\!=\!3/2$ spin-orbit
quartet.
With orbitals optimized for an average of $5d^5$ states, i.e., $^2T_{2g}$ ($t_{2g}^5$),
$^4T_{1g}$ ($t_{2g}^4e_g^1$), $^4T_{2g}$ ($t_{2g}^4e_g^1$) and $^6\!A_{1g}$ ($t_{2g}^3e_g^2$),
the $j\!=\!3/2$-like components lie at 0.82 and 0.86 eV above the $j\!\approx\!1/2$ doublet,
by MRCI+SOC computations (see Table\,\ref{dd_exc}).
If the reference active space in the prior CAS self-consistent-field (CASSCF) calculation 
\cite{book_QC_00} is restricted to only three ($t_{2g}$) orbitals and five electrons, the
relative energies of the $j\!\approx\!3/2$ components in the subsequent MRCI+SOC treatment are
somewhat lower, 0.69 and 0.73 eV.   
The Ir $t_{2g}$ to $e_g$ transitions require excitation energies of at least 3 eV according to
the MRCI data in Table\,\ref{dd_exc}, similar to values computed for $\alpha$-Li$_2$IrO$_3$ \cite{Ir213_rixs_gretarsson_2013}.

While the quantum chemistry results for the on-site excitations in $\beta$-Li$_2$IrO$_3$ resemble
very much the data for the quasi-2D honeycomb iridates, the computed intersite effective interactions
show significant differences.
The latter were estimated by MRCI+SOC calculations for embedded fragments having two edge-sharing IrO$_6$
octahedra in the active region.
As detailed in earlier work \cite{Na2IrO3_vmk_14,Li2IrO3_vmk_14,Ba214_vmk_14}, the {\it ab initio}
quantum chemistry data for the lowest four spin-orbit states describing the magnetic spectrum of two
NN octahedra is mapped in our scheme onto an effective spin Hamiltonian including both isotropic 
Heisenberg exchange and symmetric anisotropies.
Yet the spin-orbit calculations, CASSCF or MRCI, incorporate all nine triplet and nine singlet
states that arise from the two-Ir-site $t_{2g}^5$--$t_{2g}^5$ configuration (see SI).
The MRCI treatment includes the Ir $5d$ electrons and the O $2p$ electrons at the two bridging ligand
sites.

MRCI+SOC results for the NN effective couplings are listed in Table\,II.
The two, structurally different sets of Ir-Ir links are labeled {\it B1} and {\it B2}, see Fig.\,1.
For each of those, the O ions are distributed around the Ir sites such that the Ir-O-Ir bond angles
deviate significantly from 90$^{\circ}$.
While the {\it B1} links display effective $D_2$ point-group symmetry
\footnote{
The effective symmetry of a block of two NN octahedra is dictated not only by the precise arrangement of the
O ions coordinating the two magnetically active Ir sites but also by the symmetry of the extended
surroundings.},
the {\it B2} bonds possess $C_i$ symmetry, slightly away from $C_{2h}$ due to small differences
between the Ir-O bond lengths on the Ir$_2$O$_2$ plaquette of two Ir ions and two bridging ligands
(2.025 vs 2.023 \AA \ \cite{Takayama14}).
The absence of an inversion center allows a nonzero antisymmetric exchange on the {\it B1} links.
However, our analysis shows this antisymmetric Dzyaloshinskii-Moriya coupling is the smallest effective
parameter in the problem --- two orders of magnitude smaller than the dominant NN interactions, i.e.,
the Kitaev exchange.
On this basis and further symmetry considerations (see the discussion in
Refs.\,[\citeonline{Na2IrO3_vmk_14,Li2IrO3_vmk_14,Ba214_vmk_14,Sr214_niko_15}]), 
the effective spin Hamiltonian for the {\it B1} links is assumed $D_{2h}$-like and in the local
Kitaev reference frame (with the $z$ axis perpendicular to the Ir$_2$O$_2$ plaquette and $x$, $y$
within the plane of the plaquette~\cite{Ir213_KH_jackeli_09,Na2IrO3_vmk_14}) it reads
\begin{equation}
  \mc{H}_{ij}^{\mathit{B1}} = J\, \tilde{\vec{S}}_i\!\cdot\!\tilde{\vec{S}}_j
  + K\,{\tilde S}_i^z {\tilde S}_j^z 
  +\Gamma_{xy}\,(\tilde{S}_i^x\tilde{S}_j^y+\tilde{S}_i^y\tilde{S}_j^x)\;,
  \label{eqn:d2h}
\end{equation}
where $\tilde{\bf  S}_i$ and $\tilde{\bf S}_j$ are pseudospin 1/2 operators, $K$ defines the Kitaev
component and $\Gamma_{xy}$ is the only non-zero off-diagonal coupling of the symmetric anisotropic
tensor. 

For the {\it B2} units of edge-sharing IrO$_6$ octahedra, the effective spin Hamiltonian reads
in the local Kitaev coordinate frame as
\begin{equation}
  {\mathcal H}_{ij}^{\mathit{B2}} = J\,{\tilde {\bf S}}_i\cdot {\tilde {\bf S}}_j
+ K\,{\tilde S}_i^z {\tilde S}_j^z +
\displaystyle\sum\limits_{\alpha\ne\beta} 
\Gamma_{\alpha\beta}({\tilde S}_i^{\alpha}{\tilde S}_j^{\beta} + {\tilde S}_i^{\beta}{\tilde S}_j^{\alpha})\;.
\label{eqn:c2h}
\end{equation}
We find for the {\it B2} links that slight distortions lowering the bond symmetry from $C_{2h}$ to $C_i$
have minor effects on the computed wave functions and the quantum chemistry data can be safely mapped
onto a $C_{2h}$ model.
For $C_{2h}$ symmetry, the elements of the symmetric anisotropic tensor are such that
$\Gamma_{zx}\!=\!-\Gamma_{yz}$.

The wave functions for the low-lying four states in the two-Ir-site problem can be conveniently expressed
in terms of 1/2 pseudospins as in Table\,II.
In $D_{2}$ symmetry ({\it B1} links) these pseudospin wave functions, singlet $\Phi_{\mathrm{S}}$ 
and triplet $\Phi_1$, $\Phi_2$, $\Phi_3$, transform according to the $A_u$, $B_2$, $B_1$ and $A_u$
irreducible representations, respectively.
For (nearly) $C_{2h}$ symmetry ({\it B2} links), $\Phi_{\mathrm{S}}$, $\Phi_1$, $\Phi_2$ and $\Phi_3$ 
transform according to $A_g$, $B_u$, $B_u$ and $A_u$, respectively.
The amount of $\Phi_{\mathrm{S}}$--$\Phi_3$ ({\it B1}) and $\Phi_1$--$\Phi_2$ ({\it B2}) mixing 
(see Table\,II) is determined by analysis of the ``full" spin-orbit wave functions obtained in the
quantum chemistry calculations.

As seen in Table\,II, for each set of Ir-Ir links in $\beta$-Li$_2$IrO$_3$, {\it B1} and
{\it B2}, both $J$ and $K$ are FM.
In contrast, $J$ is AF for all pairs of Ir NN's in honeycomb Na$_2$IrO$_3$ \cite{Na2IrO3_vmk_14} 
and features different signs for the two types of Ir-Ir links in $\alpha$-Li$_2$IrO$_3$
\cite{Li2IrO3_vmk_14}.
The Kitaev exchange, on the other hand, is found to be large and FM in all 213
compounds, see Table\,II and Refs.\,\citeonline{Na2IrO3_vmk_14,Li2IrO3_vmk_14}.
In addition to the Kitaev coupling, sizable off-diagonal symmetric anisotropic interactions
are predicted.
In $\beta$-Li$_2$IrO$_3$, these are FM for the {\it B1} bonds and show up with both $+$ and $-$
signs for the {\it B2} links \footnote{The sign of these terms is with respect to the local Kitaev reference frame.},
see Table\,II.

\subsection{Magnetic Phase Diagram}
\begin{figure}[!b]
\includegraphics[width=0.40\linewidth,scale=0.8]{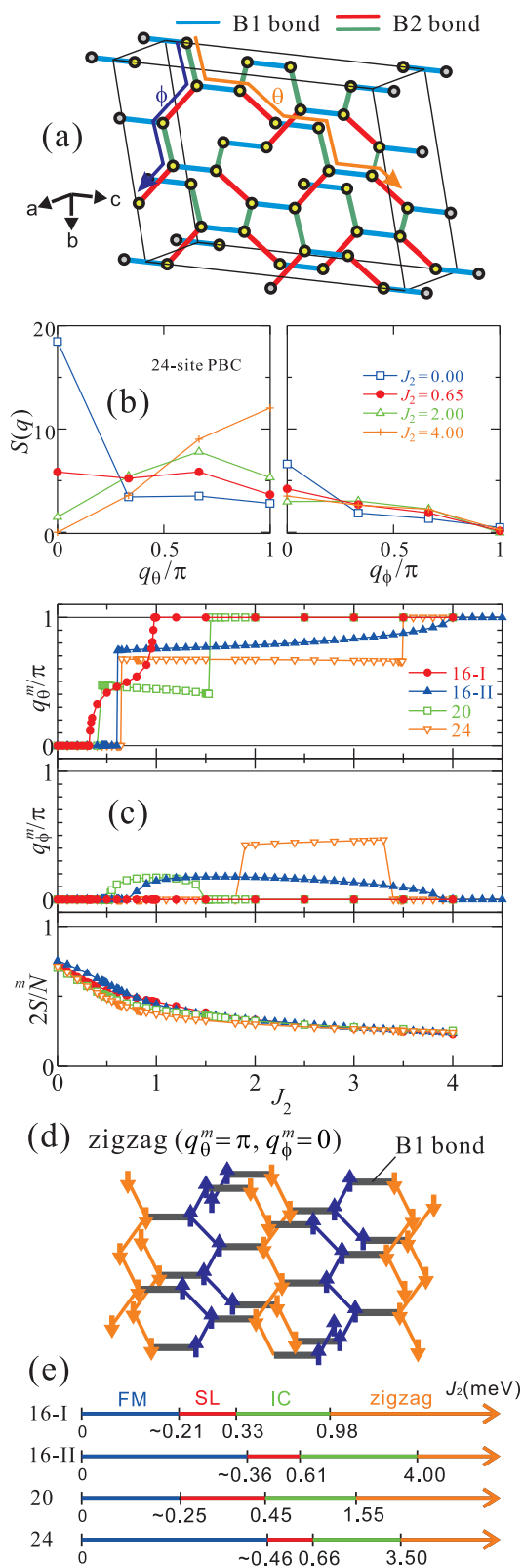}
\caption{
(a) Sketch of the 24-site ``periodic'' cluster.
(b) Static spin-structure factor along paths $\theta$ and $\phi$, see text.
(c) Propagation vectors $q_\theta^m$, $q_\phi^m$ and total spin $2S/N$ for the periodic clusters, as functions of $J_2$.
(d) AF zigzag order on the hyper-honeycomb lattice.
(e) Magnetic phase diagrams obtained for the periodic clusters.
}
\label{ED_figs1}
\end{figure}

Having established the nature and the magnitude of the NN effective spin couplings, we now turn
to the magnetic phase diagram of $\beta$-Li$_2$IrO$_3$.
In addition to the NN MRCI+SOC data of Table\,II, we have to take into account explicitly the 
second-neighbor Heisenberg interactions.
Due to the 3D nature of the iridium lattice, with alternate rotation of two adjacent {\it B2} bonds
around the {\it B1} link with which both share an Ir ion, one can safely assume that the
third-neighbor exchange is vanishingly small.
Results of ED calculations for an extended (pseudo)spin Hamiltonian including the MRCI
NN interactions and a variable second-neighbor Heisenberg coupling parameter $J_2$
are shown in Fig.\,2.
Different types of clusters were considered, with either 16, 20 or 24 Ir sites.
The 24-site cluster used in ED calculations with periodic boundary conditions is displayed in
Fig.~\ref{ED_figs1}(a) while the structure of the smaller clusters is detailed in SI.


In order to investigate the magnetic properties of $\beta$-Li$_2$IrO$_3$, we calculated the
static spin-structure factor
$S(q)\!=\!\sum_{ij} \langle {\tilde {\bf S}}_i\!\cdot\!{\tilde {\bf S}}_j \rangle \exp[i q(r_i\!-\!r_j)]$ 
along two paths denoted as $\theta$ ($bc$-diagonal) and $\phi$ ($ab$-diagonal) in 
Fig.~\ref{ED_figs1}(a), where the distance between neighboring {\it B1} bonds is taken as 1. 
The results for several $J_2$ values with the 24-site cluster are plotted in Fig.~\ref{ED_figs1}(b).
The propagation vector for each path ($q_{\theta}^m$/$q_{\phi}^m$), determined as 
the wave number $q$ providing a maximum of $S(q)$, is plotted in Fig.~\ref{ED_figs1}(c). 
For $J_2\!=\!0$ the ground state is characterized by long-range FM order, i.e., 
$q_\theta^m\!=\!q_\phi^m\!=\!0$, consistent with a previous classical Monte Carlo study
\cite{lee_CMC,Lee_HH_magphases_2015}.
Given the strong FM character of the NN exchange, ground states different from FM order
are only obtained for finite AF $J_2$.
With increasing strength of the AF $J_2$, $q_\theta$ develops finite values starting
at $J_2\!=\!J_{2,c1}$ and reaches $\pi$ at $J_2\!=\!J_{2,c2}$ whereas $q_\phi$ is
finite but small in the range $J_{2,c1}\!<\!J_2\!<\!J_{2,c2}$ and zero otherwise.
This evidences two magnetic phase transitions, from FM to IC order and
further to a commensurate ground state.
The latter commensurate structure corresponds to zigzag AF order, a schematic picture of which 
is shown in Fig.~\ref{ED_figs1}(d).
The ED results for the four different types of periodic clusters are here in good overall
agreement, as shown in Fig.~\ref{ED_figs1}(e).
Some differences arise only with respect to the precise position of the critical points.

An intriguing feature is the appearance of a SL state in between the FM and IC phases. 
Since the total spin $2S/N$ falls off rapidly and continuously near $J_2\!=\!J_{2,c1}$ [see
Fig.~\ref{ED_figs1}(c)], the FM ground state is expected to change into SL before reaching
the IC regime. It can be confirmed by a structureless static spin-structure factor, like
nearly flat $q$-dependence of $S(q)$ at $J_2=0.65$ in Fig.~\ref{ED_figs1}(b).
In Fig.~\ref{ED_figs1}(e) we also provide the critical values marking the transition between the FM and SL states.
This was estimated as the point where any of the $\langle \tilde{\vec{S}}_i\!\cdot\!\tilde{\vec{S}}_j \rangle\!$
expectation values turn negative, which implies a collapse of long-range FM order. 
Importantly, we find that the SL phase shows up in each of the four different types of periodic clusters. 
A more detailed analysis of the spin-spin correlations is provided in SI.

\section{Discussion}

\begin{figure}[!b]
\includegraphics[width=0.40\linewidth,scale=0.8]{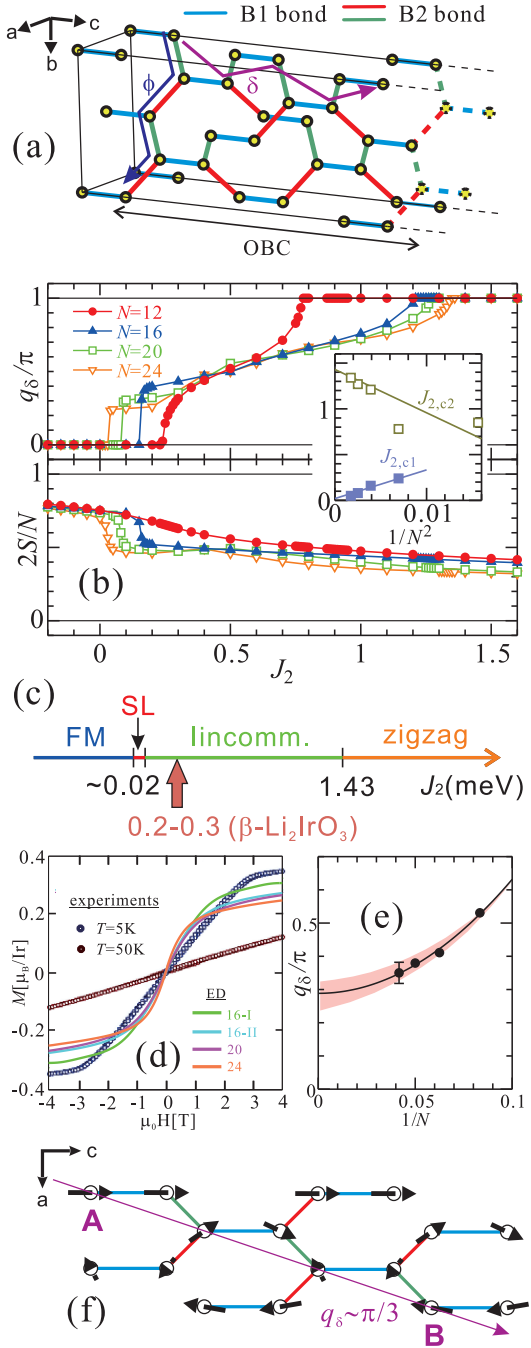}
\caption{
(a) Sketch of the cluster with open boundaries along the $c$ direction.
(b) Propagation vector $q_\delta^m$ and total spin $2S/N$ for our ``open'' clusters, as function of $J_2$.
Inset: finite-scaling analysis of the critical points.
(c) Magnetic phase diagrams obtained by ED.
(d) Experimental (see Ref.\,\citeonline{Takayama14}) and theoretical magnetization
curves for $\beta$-Li$_2$IrO$_3$.
The latter are obtained with either $J_2$=0.2 (periodic 16- and 20-site clusters) or
$J_2$=0.3 meV (periodic 24-site cluster) and the NN MRCI couplings from Table\,II.
(e) Finite-scaling analysis of $q_\delta^m$ at $J_2$=0.3 meV using the open clusters.
(f) Experimental results of the magnetic structure for $\beta$-Li$_2$IrO$_3$ (see Ref.\,\citeonline{biffin_xray}).
}
\label{ED_figs2}
\end{figure}

Typically, a commensurate-to-IC transition critical point tends to be overestimated by using periodicity.
For estimating more precisely the critical $J_2$ values we therefore additionally studied
clusters
   with open boundary conditions along the $c$ direction.
Also, for a direct comparison between our ED results and the experimentally observed magnetic
structure, we introduce an additional path $\delta$ ($ac$-diagonal), sketched in Fig.~\ref{ED_figs2}(a).
The size of the cluster along $a$ and $b$ has insignificant effect on the computed critical
$J_2$ values because $q_\phi^m$ ($ab$-diagonal) is either zero, around the critical points
(periodic 24-site cluster), or very small, in the IC phase (periodic 16- and 20-site clusters), 
as seen in Fig.~\ref{ED_figs1}(c).

The value of the propagation vector along the $\delta$-path ($q_\delta^m$) is shown in
Fig.~\ref{ED_figs2}(b) as function of $J_2$ for various cluster ``lengths'' in the $c$
direction.
The inset displays a finite-size scaling analysis for the critical values.
In the infinite-length limit, we find $J_{2,c1}\!=\!0.02$ and $J_{2,c2}\!=\!1.43$ meV.
The corresponding phase diagram is provided in Fig.~\ref{ED_figs2}(c).
Similar critical points, i.e., $J_{2,c1}\!=\!0.02$ and $J_{2,c2}\!=\!1.48$ meV, are
obtained for $q_\theta^m$ (see SI).

As shown in Fig.~\ref{ED_figs2}(b), the dropdown of $2S/N$ near $J_2\!=\!J_{2,c1}$ is
more clearly seen than in the case of periodic clusters because the formation of IC order is
not hindereded for open clusters. Defining the FM-SL $J_{2,c1}$ critical value as the point where 
$\langle \tilde{\vec{S}}_i\!\cdot\!\tilde{\vec{S}}_j \rangle$ turns negative for any 
($i$,$j$) pair, the SL phase in the vicinity of $J_2\!\approx\!J_{2,c1}\!=\!0.02$ meV would have
a width of about $0.01J_2$.
In other words, a very tiny FM $J_2$ coupling may drive the system from FM order to a SL state.
With further increasing $J_2$, the system goes through an IC phase to AF
zigzag order at $J_2\!=\!1.43$ meV.

To finally determine the value of $J_2$ in $\beta$-Li$_2$IrO$_3$, we fitted the 
magnetization curve obtained by ED calculations at $T\!=\!0$\,K [see Fig.~\ref{ED_figs2}(d)] 
to the experimental data at $T\!=\!5$\,K \cite{Takayama14}. 
Such an exercise yields $J_2\!=\!0.2\!-\!0.3$ meV, i.e., $J_2 \approx 0.1 J_{2,c2}$, so that
the system is relatively far from the instability to zigzag order but very close to the transition
to the SL ground state.
Since with increasing $J_2$ the propagation vector $q_\delta^m$ of the IC phase
increases smoothly from that of the SL ($q_\delta^m\!=\!0$) to that of the zigzag state
($q_\delta^m=\pi$), long-wavelength IC order with a small propagation vector is expected
for $\beta$-Li$_2$IrO$_3$. By performing a finite-size scaling analysis of $q_\delta^m$ at
$J_2\!=\!J_{2,c1}(N)+0.28$ meV, we obtain $q_\delta^m/\pi\!=\!0.28 \pm 0.04$ for $J_2\!=\!0.3$ meV
in the infinite-length limit.
An experiment-based estimate for $q_\delta^m$ can be extracted from recent magnetic resonant
x-ray diffraction data \cite{biffin_xray} [see Fig.~\ref{ED_figs2}(f)]; the spins on sites A and
B (their distance is three lattice spacings) have almost opposite directions, which leads to
$q_\delta^m/\pi\!\sim\!1/3$.
That fits reasonably well our theoretical estimate.
The stabilization of an IC state by $J_2$ couplings has been previously discussed for 1D zigzag
chains like the path we label here as $\delta$ in Ref.\,\citeonline{Kimchi_HH_unified_2015}.

The value extracted for $J_2$ from our fit of the magnetization data is thus within our theoretical framework fully consistent with the 
experimentally observed IC magnetic order in $\beta$-Li$_2$IrO$_3$. Nevertheless we find that the system is remarkably 
close to a three-dimensional spin-liquid ground state, which can be reached by a minute change of $\sim$ 0.25 meV, an energy scale that corresponds to about 3K, in the second-neighbour 
exchange parameter $J_2$. Changes of this order of magnitude can easily be induced by pressure or strain.

\section{Acknowledgments}
We thank N.~Bogdanov for helpful discussions.
LH and SN acknowledges financial support from the German Research Foundation (Deutsche
Forschungsgemeinschaft, DFG --- SFB 1143 and HO4427). JvdB acknowledges support from the Harvard-MIT CUA. 
Part of the calculations have been performed using the facilities of the
Center for Information Services and High Performance Computing (ZIH) of the
Technical University of Dresden.


\begin{table}[!ht]
\caption{
Ir$^{4+}$
5$d^5$ multiplet structure in $\beta$-Li$_2$IrO$_3$, all numbers in eV.
Due to the noncubic environment, the $T_{2g}$/$T_{1g}$ (and spin-orbit coupled $j\!=\!3/2$)
states are split appart.
We still use however notations corresponding to $O_h$ symmetry.
Only the lowest and highest Kramer's doublets are shown for each set of higher-lying spin-orbit states.
}
\label{dd_exc}
\begin{tabular}{lll}
\hline
\hline\\[-0.3cm]
States             &MRCI                            &MRCI+SOC ($\times$2)            \\
\hline\\[-0.3cm]
$t_{2g}^5$         &0,    0.07, 0.11 ($^2T_{2g}$)   &0         \  ($j\!\approx\!1/2$)\\
&                                                   &0.82, 0.86\, ($j\!\approx\!3/2$)\\[0.1cm]

$t_{2g}^4e_g^1$    &2.99, 3.01, 3.02 ($^4T_{1g}$)   &3.32, \dots 3.79                \\
                   &3.60, 3.65, 3.66 ($^4T_{2g}$)   &4.23, \dots 4.50                \\[0.1cm]

$t_{2g}^3e_g^2$    &5.01             ($^6\!A_{1g}$) &5.87, \dots 5.87                \\
\hline
\hline
\end{tabular}
\end{table}

\begin{table}[!ht]
\caption{
MRCI splittings among the four low-lying magnetic states and effective exchange couplings (meV)
for two NN IrO$_6$ octahedra in $\beta$-Li$_2$IrO$_3$.
A {\it local} coordinate frame is used for each Ir-Ir link ($x$ along the Ir-Ir bond,
$z$ perpendicular to the Ir$_2$O$_2$ plaquette).
For {\it B1} bonds, the weight of $\Phi_{\rm S}$ in $\Psi_{\mathrm{S}}$ and of
                                  $\Phi_{\rm 3}$ in $\Psi_{\mathrm{3}}$ is $\approx$99\%.
For {\it B2} links, the $\Phi_{\mathrm{1}}$--$\Phi_{\mathrm{2}}$ mixing is approximately
3\%--97\%, where
$\Phi_{\mathrm{1}}= (\uparrow\downarrow   +\downarrow\uparrow)/\sqrt2$,
$\Phi_{\mathrm{2}}= (\uparrow\uparrow +\uparrow\uparrow)/\sqrt2$,
$\Phi_{\mathrm{3}}=(\uparrow\uparrow   -\uparrow\uparrow)/\sqrt2$
and
$\Phi_{\mathrm{S}}=(\uparrow\downarrow   -\downarrow\uparrow)/\sqrt2$,
see text.
}
\label{tab:magcoup}
\begin{tabular}{crr}
\hline
\hline\\[-0.25cm]
Energies \& effective couplings
&{\it B1}\footnotesize{$^1$}\,
&{\it B2}\footnotesize{$^2$}\,
\\
\hline\\[-0.15cm]
$E_2$ ($\Psi_{\mathrm{2}}$)            &$0.0$     &$0.0$ \\
$E_3$ ($\Psi_{\mathrm{3}}$)            &$2.1$     &$4.2$ \\
$E_1$ ($\Psi_{\mathrm{1}}$)            &$8.4$     &$8.3$ \\
$E_{\mathrm{S}}$ ($\Psi_{\mathrm{S}}$) &$8.7$     &$10.5$\\[0.20cm]
$J$                                    &$-0.3$    &$-2.4$ \\
$K$                                    &$-14.7$   &$-11.7$\\
$\Gamma_{xy}$                          &$-2.1$    &$-3.9$ \\
$\Gamma_{zx}\!=\!-\Gamma_{yz}$         &---       &$ 2.0$ \\
\hline
\hline
\end{tabular}\\
\footnotesize{$^1$ $\measuredangle$(Ir-O-Ir)=94.7$^{\circ}$, $d$(Ir-Ir)=2.98, $d$(Ir-O$_{1,2}$)=2.025 \AA \ \cite{Takayama14}.}\\
\footnotesize{$^2$ $\measuredangle$(Ir-O-Ir)=94.4$^{\circ}$, $d$(Ir-Ir)=2.97, $d$(Ir-O$_1$)=2.025, $d$(Ir-O$_2$)=2.023 \AA \ \cite{Takayama14}. O$_1$ and O$_2$ are the two bridging O's.}
\end{table}

\end{document}